# Functionalized MXenes as Effective Polyselenides Immobilizer for Lithium-Selenium Batteries: A Density Functional Theory (DFT) Study

Rahul Jayan, Md Mahbubul Islam*

Department of Mechanical Engineering, Wayne State University, Detroit, MI 48202

*Corresponding author: mahbub.islam@wayne.edu

The practical applications of lithium selenium (Li-Se) batteries are impeded primarily due to the dissolution and migration of higher order polyselenides ($Li_2Se_n$) into the electrolyte (known as *shuttle effect*) and inactive deposition of lower order polyselenides. The high electrical conductivity and mechanical strengths of MXenes make them a suitable candidate to provide adequate anchoring to prevent polyselenides dissolution and improved electrochemical performance. Herein, we used density functional theory (DFT) calculations to understand the binding mechanism of $Li_2Se_n$ on graphene and surface functionalized $Ti_3C_2$ MXenes. We used graphene as reference material to assess $Li_2Se_n$ binding strengths on functionalized $Ti_3C_2X_2$ (where X = S, O, F, and Cl). We observed that $Ti_3C_2S_2$ and $Ti_3C_2O_2$ exhibit superior anchoring behavior compared to graphene, $Ti_3C_2F_2$, and $Ti_3C_2Cl_2$. The calculated $Li_2Se_n$ adsorption strength provided by S and O terminated $Ti_3C_2$ are stronger than the commonly used ether-based electrolyte, which is a requisite for effective suppression of the $Li_2Se_n$ shuttling. The adsorbed $Li_2Se_n$ on $Ti_3C_2X_2$ and graphene retains their structural integrity without a chemical decomposition. The density of states (DOS) analysis exhibits that the conductive behavior of the $Ti_3C_2X_2$ is preserved even after $Li_2Se_n$ adsorption, which can stimulate the electrochemical activity of involved $Li_2Se_n$ chemistry. Based on our unprecedented results, $Ti_3C_2S_2$ and $Ti_3C_2O_2$ are found to exhibit superior anchoring behavior for $Li_2Se_n$ adsorption, which can be leveraged for designing effective selenium-based cathode materials to boost the electrochemical performance of the Li-Se battery system.

## 1. Introduction

The lithium ion batteries (LIBs) as approaching their theoretical limits, alternative battery technologies are being developed aiming large scale applications such as electric vehicle (EV). The implementation of the rechargeable batteries in EVs requires high energy density, longer cycle life, lower cost, and higher efficiency.[1] The wide abundance and astounding energy density of chalcogen group elements such as O, S, and Se make them suitable candidates for future energy storage applications. However, owing to the electrolyte decomposition and the need for excess Li in the anode, the commercialization of the $Li-O_2$ batteries are far apart from reality.[2] However, rechargeable lithium sulfur (Li-S) batteries can serve the purpose because of its high theoretical energy density (2600 W h $kg^{-1}$) and capacity (1673 mA h $g^{-1}$) values approximately five times higher when compared to the commercial LIBs, and they are cost effective, widely abundant, and low in toxicity.[3,4] However, practical realization of Li-S batteries is thwarted due to various challenges such as the dissolution of intermediate polysulfides into the electrolytes, commonly known as *shuttle effect,* and the less utilization of the active materials due to the formation of insoluble polysulfides.[3,5] Although strategies such as introducing conductive porous matrices,[6] additives,[7] and usage of suitable electrolytes[8] has been established into the Li-S batteries; the practical applications is still restricted which warrant the search of other substitutes for Li-S battery cathode materials to deliver high energy density and specific capacity and meet the requirements of the next generation energy landscape.

Rechargeable lithium selenium (Li-Se) batteries are analogous to Li-S counterpart in terms of chemical properties as the selenium reacts with the lithium ions to form lithium polyselenides ($Li_2Se_n$). Although the gravimetric capacity of Se (678 mA h $g^{-1}$) is lesser when compared to the S (1672 mA h $g^{-1}$), but its higher volumetric capacity (3253 mA h $cm^{-3}$) is comparable to S (3467 mA h $cm^{-3}$).[9,10] Moreover, the electronic conductivity of Se (1x $10^{-3}$ S/m) is at least twenty orders of magnitude higher than the S (5 x $10^{-28}$ S/m) that triggers better electrochemical activities of Se.[8] The Li-Se batteries is a prospective candidate to meet the requirements of the higher volumetric energy density for large scale applications as an effective alternative to Li-S batteries. However, the dissolution of soluble intermediate $Li_2Se_n$ into the electrolytes ensue a low battery capacity and limits the cycle performance of Li-Se batteries.[11,12] The carbonate-based electrolytes have been reported to suppress the polyselenides shuttling due to a single step conversion to $Li_2Se$; however, the performance improvement is quite limited.[13] The nucleophilic nature of Se can cause reactions with the carbonate-based solvents, thus detrimentally affects the battery capacity and cyclability.[14] Furthermore, the large volumetric change of Se cathodes during charge/discharge cycles results in fracture and pulverization of the Se active materials.[13]

The Se confinement into carbon mesopores are reported to provide the space required to accommodate the volume changes and suppress intermediate polyselenides shuttle during the lithiation/delithiation cycles.[15–20] Liu and coworkers incorporated Se molecules into the microporous carbon nanofibers via vacuum and heat treatment.[21] This process leads to a better anchoring and homogenous distribution of the selenides on the carbon matrix.[21] Abouimrame et al.[9] has studied the Se and Se-S composite cathode materials and reported that the high utilization of the bulk Se (ca. 45 wt. %) active material upon extended cycling cannot be achieved in S bulk cathode indicating the better activity and lesser shuttling of Se. The stable capacity of 415 mA h g$^{-1}$ at 0.2 C for 100 cycles was realized through homogeneously distributed Se on polypyrole carbonized and KOH activated interconnected porous carbon nanofibers.[22] The CoSe$_2$-porous carbon composites,[12] graphene-selenium hybrid microballs,[23] hollow double-shell Se@CNx nanobelts,[11] and Ag$_2$Se coating onto the Se encapsulates, and various metal organic frameworks are reported to alleviate but not to entirely eliminate the dissolution of polyselenides as required to achieve the target performance of Li-Se batteries. The Se incorporated into the 3D interconnected porous carbon nanofibers are investigated as possible means to suppress the active material dissolution via the uniform Se distribution onto confined mesopores, which readily acts as a cushion for volume expansion and enhances the electrochemical performance.[24] Moreover, the polyselenides migration can be suppressed by constraining Se into microporous N-doped carbon.[25] The hierarchical architecture developed by the fusion of graphene and 3D porous carbon nanoparticles are probed to improve the electric and ionic conductivities and suppress the dissolution of polyselenides.[26] He at al.[27] introduced 3D graphene - CNT/Se cathode and found that the conductive network provided organized channels for the Li diffusion and electron transfer and further the hierarchical structure prevented the polyselenides dissolution into electrolytes. Although selenium confinement onto the carbon matrix could reduce the migration, the optimization of the micropore size in carbon structure is a challenging task. As the pore size of the carbon matrix depends on the overall architecture, the minimal size could originate difficulties in accommodating the electroactive material (Se) and diffusion of Li ions.[28] Since the larger micropores could promote the dissolution of Se upon charging and drastically affect the electrochemical performance,[13] which in turn prompts further research efforts to investigate alternate anchoring materials (AMs) for the Li-Se system.

Researchers have sought various novel materials and strategies to address the polyselenides shuttling during the charge/discharge process. Recently, MXenes, a class of 2-D materials, grabbed special attention because of its multiple attractive features such as high electronic conductivity, high surface area, and catalytically active sites for facilitating the Li$_2$Se$_n$ conversion reactions.[29] However, the bare MXenes cannot be directly used in the chalcogenides system due to the high reactivity of polychalcogenides towards transition metals that causes chemical decomposition of polychalcogenides.[30,31] Thus, surface functionalization are required, and commonly chalcogenides and halogens groups are employed.[30,31,32–35] MXenes have exhibited promising performances in Li-S batteries. Wang et al.[36] studied the shuttling effect of polysulfides in Li-S batteries using density functional theory (DFT) calculations with the anchoring materials such as bare V$_2$C, V$_2$CO$_2$, and V$_2$CS$_2$ and reported that the sulfur functionalized MXenes exhibits adequate thermal and chemical stability and polysulfides binding strength. The study claimed that the material lowers the barrier for Li$_2$S decomposition and lithium diffusion, and retained the metallic properties even after the adsorption of polysulfides.[36] Furthermore, Ti$_3$C$_2$ MXene functionalized with N, O, F, S, and Cl has been investigated as cathode materials for Li-S batteries and O and S surface functionalization are found to provide strong binding behavior with the polysulfides and superior catalytic activities for the conversion of polysulfides.[29] However, the role of MXenes in the context of polyselenides retention has not been studied as a plausible route to improve the capacity and cycle performance of Li-Se batteries.

The excellent electronic properties of MXenes and their promising performance in Li-S batteries can be leveraged for designing MXenes based Se cathode materials. To the best of our knowledge, no studies have been conducted to elucidate the polyselenides binding characteristics and mechanisms on bare and functionalized MXenes. Herein, we perform first-principles based

DFT calculations to understand the structural stability, adsorption behavior, and electronic properties of the MXenes and graphene. We use graphene as reference material to compare the anchoring performance of Mxenes. We start with bare $Ti_3C_2$ MXenes and investigate the polyselenides binding behavior and found that bare MXenes causes strong binding and decompositions of polyselenides. This observation motivates us to incorporate four disparate surface functionalization and extensively examine their role on the underlying characteristics and the mechanisms of polyselenides anchoring on MXenes as host materials for the cathodes of Li-Se batteries.

## 2. Calculation methods

All the first-principles calculations were executed using the Vienna ab initio simulation package (VASP).[37] We used Projector Augmented Wave (PAW) pseudopotential to investigate the electron-ion interaction and Perdew-Bruke-Ernzerhof (PBE) functional with the Generalized Gradient Approximations (GGA) to describe the electron-electron exchange correlations.[38] The kinetic energy cut off was selected as 520 eV for the plane wave basis calculations. A vacuum space of 30 Å was added to avoid the interactions across the periodic boundary. We accounted for van der Waal interactions using the DFT-D3 method to provide better accuracy in evaluating the binding strength of polyselenides with the MXenes.[39,40] The conjugate gradient method was employed for geometry optimization, and the convergence thresholds of energy and force are fixed to be $10^{-4}$ eV and 0.025 eV/Å, respectively. The calculations were performed with 4 x 4 supercell of $Ti_3C_2$. The Brillouin zone sampling for the geometric relaxations and the electronic structure calculations were implemented using the 5 x 5 x 1 and 11 x 11 x 1 k-points mesh generated by the Monkhorst-Pack grid scheme, respectively. The Bader charge analysis[41] was performed to study the charge transfer between the AMs and $Li_2Se_n$ and the charge density difference was calculated using the equation

$$\rho_b = \rho_{adsorbed\ state} - (\rho_{adsorbent} + \rho_{AM})$$

Where $\rho_{adsorbed\ state}$, $\rho_{adsorbent}$, and $\rho_{AM}$ refers to the charge transfer of $Li_2Se_n$ adsorbed AM, the isolated $Li_2Se_n$ and the AM, respectively. The charge density difference data were visualized using VESTA code.[42] The electronic behavior of bare and surface functionalized MXenes was probed using the density of states (DOS) analysis.

## 3. Results and discussion

**$Li_2Se_n$ Structures**. The underlying chemistry in Li-Se and Li-S batteries are quite analogous. During the discharge process, the lithium ions released from the anode reacts with the Se cathode to generate intermediate polyselenides leading to the final discharge products such as $Li_2Se_2$ and $Li_2Se$. The most stable structures of $Li_2Se_n$ are obtained via relaxation simulations, and the optimized $Li_2Se_n$ structures are shown in Figure 1. The observed three-dimensional geometries of $Li_2Se_n$ molecules arises due to the tendency of the $Li^+$ cations to chemically bond with electron rich terminal Se atoms. The puckered ring structured $Se_8$ possess $D_{4d}$ symmetry and the shortest Se-Se bond length is 2.353 Å. The higher ordered soluble polyselenides such as $Li_2Se_8$, $Li_2Se_6$, and $Li_2Se_4$ exhibit $C_2$ symmetry with the minimum Li-Se bond distances of 2.542, 2.581, and 2.524 Å, respectively. The insoluble $Li_2Se_2$ and $Li_2Se$ exhibit $C_{2v}$ symmetry and the corresponding shortest bond lengths are 2.387 and 2.238 Å. We observe a decreasing trend in the Li-Se bond lengths from higher to lower order polyselenides, and the behavior can be ascribed to the stronger Li-Se covalent interactions.

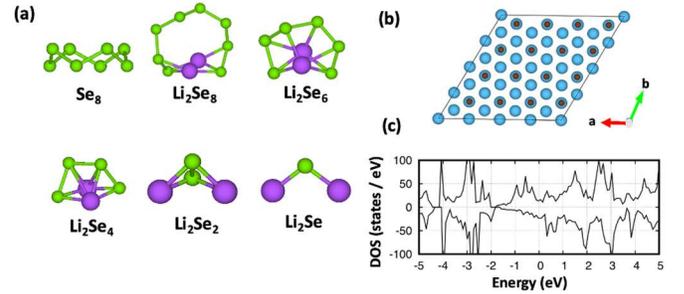

Figure 1. (a) The optimized geometric structures of $Li_2Se_n$ (b) side view of bare $Ti_3C_2$ (c) Total density of states (DOS) of $Ti_3C_2$. Color codes: Green: Selenium, Purple: Lithium, Blue: Titanium, and Gray: Carbon. The partial DOS of $Ti_3C_2$ is provided in Figure S1 of the supplementary information (SI)

**Graphene.** Carbon-based selenium composite materials are widely used in Li-Se batteries.[17,24–26] As a representative model carbon material, we considered graphene to compare the $Li_2Se_n$ anchoring performance of MXenes. We aim to understand the structural stability and binding strength of $Li_2Se_n$ adsorbed on graphene. The Graphene monolayer used for our calculations contains a 5 x 5 supercell, and the lattice parameter is 12.338 Å. The $Li_2Se_n$ molecules were allowed to adsorb on the graphene by placing the two Li atoms on top of C atoms. We performed simulations to assess the anchoring behavior of the $Li_2Se_n$ on graphene, and the adsorption energies are calculated using the formula

$$E_{ads} = E_{Li_2Se_n} + E_{AM} - E_{Li_2Se_n+AM}$$

Where $E_{Li2Sen}$ $E_{AM}$ and $E_{Li2Sen + AM}$ denotes the total energies of polyselenides, AM, and polyselenide adsorbed AMs, respectively. The calculated binding energies are presented in Figure 2, which compares well with other works.[43] The configurations of energetically the most favored graphene-$Li_2Se_n$ systems are displayed in Figure S2. From the figure, we could observe that the cyclooctaselenium ($Se_8$) molecule is orientated parallel to the graphene sheet with a minimum bond distance of 3.474 Å. We observe an increase in the Li-C bond length with the increasing Se concentration in the $Li_2Se_n$. The Li-C interactions retain the structural conformation of $Li_2Se_n$; however, a slight change in the average intramolecular Li-Se bond lengths are observed which are presented in Table 1. Overall, the calculated binding energies for all $Li_2Se_n$ are found to be below 0.80 eV, such weaker interactions between the $Li_2Se_n$ and graphene is indicative to the inefficient containment of the $Li_2Se_n$ within the cathode materials.

***Bare $Ti_3C_2$.*** The ineffectiveness of apolar carbon materials in preventing $Li_2Se_n$ shuttling entails the search for alternative AMs. MXenes are promising candidates to provide the required anchoring effect that can possibly stem from their polar characteristics. The chosen $Ti_3C_2$ monolayer is formulated with the three Ti atoms and two C atoms arranged in a sequence of Ti-C-Ti-C-Ti with the Ti and C atoms forming edge shared octahedral structures ($Ti_6C$).[44] The 4 x 4 $Ti_3C_2$ supercell was used in the simulations, and the relaxed lattice constant is 12.370 Å and the value is in good agreement with other reports.[29] The density of state (DOS) calculations reveal that the $Ti_3C_2$ material is electronically conductive (see Figure 1) and the projected DOS (PDOS) analysis explains that the metallic behavior arises primarily due to the 3d state of Ti as displayed in Figure S1.

The binding energy values for bare $Ti_3C_2$ adsorbed $Li_2Se_n$ are 28.72 eV for $Se_8$, 22.61 eV for $Li_2Se_8$, 11.19 eV for $Li_2Se_6$, 10.91 eV for $Li_2Se_4$, 8.23 eV for $Li_2Se_2$ and 5.49 eV for $Li_2Se$ and the optimized configurations are displayed in Figure S3. It is noted that the adsorption energy decreases while the Se count in the $Li_2Se_n$ decreases. The values indicate a strong chemical interaction between $Li_2Se_n$ and surface Ti atoms. Ostensibly, the derived absorption energies are sufficient to bind the polyselenides to prevent their dissolution into the electrolytes, however, the stronger binding results in the chemical decomposition of the intermediate polyselenides which in turn hinders reversibility of the Se cathodes. Additionally, the strong Ti-Se bonds can be readily formed during the preparation of $Ti_3C_2$/Se composites or in the course of electrochemical cycling of the Se-cathodes. The decompositions of $Li_2Se_n$ found during the adsorption precludes the use of bare $Ti_3C_2$ in selenium cathodes as AMs. Thus, in order to improve the polyselenides binding behavior with $Ti_3C_2$ and for improved performance of the battery, it is essential to implement surface modifications in the $Ti_3C_2$.

***$Ti_3C_2X_2$ (X = S, O, F, and Cl).*** To achieve adequate binding of polyselenides with the $Ti_3C_2$ MXene, we employed various surface functional groups such as X = S, O, F, and Cl. These surface terminated MXenes were selected based on their reported superior performance on thermodynamic stability, electronic properties, intercalation mechanisms, and energy storage capacity.[29,45] We obtained the optimized structures of $Ti_3C_2X_2$ via relaxation simulations with the target functional groups. The functional groups bridge between the Ti atoms at the top surface which can be observed from the optimized geometries displayed in Figure S4. The resulted surface architectures are in accordance with the other reports.[29] The lattice parameters of the surface functionalized $Ti_3C_2X_2$ have slight differences from the bare $Ti_3C_2$, and the values are tabulated in Table S1.

After obtaining the stable configurations for both $Li_2Se_n$ and the various surface functionalized $Ti_3C_2$ MXenes, we explored the interactions between them. The $Li_2Se_n$ adsorption energies on the AMs are calculated from relaxation simulations and displayed in Figure 2.

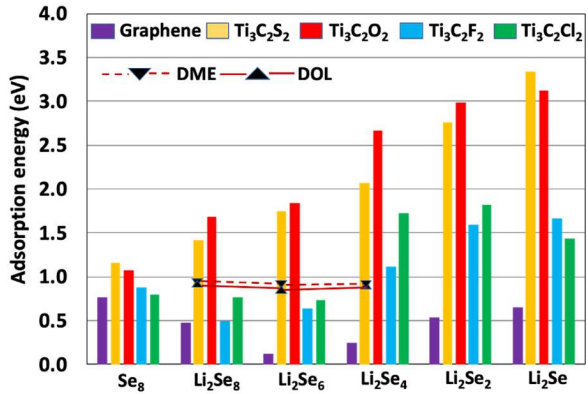

Figure 2. Adsorption energies of $Se_8$ and $Li_2Se_n$ on graphene and $Ti_3C_2X_2$ (X = S, O, F, and Cl) and the binding energies of polyselenides with electrolyte solvent molecules (DME and DOL)

For $Ti_3C_2O_2$, the relaxed structural configuration with $Li_2Se_n$ is presented in Figure 3. From the figure, one can see that the $Se_8$ molecule remains intact and parallel to $Ti_3C_2O_2$ structure without any notable deformation, and the nearest bond distance between the $Se_8$ molecule and the AM is 3.163 Å. However, we observe structural deformations in the cases of adsorbed higher order $Li_2Se_n$. From Figure 2, it can be seen that besides $Li_2Se_8$, the adsorption energy values increase with lithiation and the formation of higher to lower order polyselenides, and this trend is found the opposite to that of bare $Ti_3C_2$. Among the $Li_2Se_n$, the $Li_2Se$ (3.07eV) and $Li_2Se_6$ (1.40 eV) hold the highest and lowest binding energies, respectively. The absence of Li-atoms in the $Se_8$ molecule results in lower binding energy (1.03 eV) than the polyselenides since the binding is predominantly due to the van der Waals interactions. Furthermore, the adsorption energies of $Li_2Se_8$, $Li_2Se_4$, and $Li_2Se_2$ are calculated as 1.68 eV, 2.30 eV, and 2.86 eV, respectively. The shortest bond distance and the variation in the Li-Se bond distance for all the surface functionalized groups are presented in Table 1.

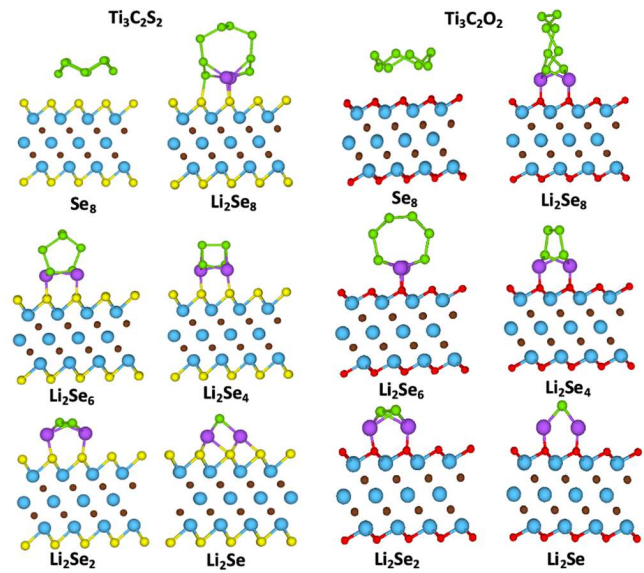

Figure 3. The side views of the most stable configurations of $Li_2Se_n$ adsorbed on $Ti_3C_2S_2$ and $Ti_3C_2O_2$. The top views are provided in Figure S5 of the SI.

Table 1. The difference in the Li-Se bond distance for the $Li_2Se_n$ adsorbed on AMs and the isolated $Li_2Se_n$ ($\Delta d_{Li-Se}$) and the minimum distance between the Li atom in $Li_2Se_n$ and the AMs ($d_{Li-AM}$)

|  |  | $Se_8$ | $Li_2Se_8$ | $Li_2Se_6$ | $Li_2Se_4$ | $Li_2Se_2$ | $Li_2Se$ |
|---|---|---|---|---|---|---|---|
| $\Delta d_{Li-Se}$ (Å) | Graphene | - | 0.000 | -0.081 | -0.025 | 0.027 | 0.071 |
|  | $Ti_3C_2S_2$ | - | 0.045 | -0.01 | 0.095 | 0.157 | 0.233 |
|  | $Ti_3C_2O_2$ | - | 0.066 | 0.038 | 0.087 | 0.157 | 0.19 |
|  | $Ti_3C_2F_2$ | - | 0.045 | -0.034 | 0.022 | 0.065 | 0.106 |
|  | $Ti_3C_2Cl_2$ | - | 0.005 | -0.075 | -0.013 | 0.027 | 0.106 |
| $d_{Li-AM}$ (Å) | Graphene | 3.474 | 2.667 | 3.580 | 3.153 | 2.535 | 2.439 |
|  | $Ti_3C_2S_2$ | 3.256 | 2.374 | 2.355 | 2.344 | 2.328 | 2.413 |
|  | $Ti_3C_2O_2$ | 3.163 | 1.867 | 1.908 | 1.864 | 1.876 | 1.866 |
|  | $Ti_3C_2F_2$ | 3.466 | 2.064 | 2.050 | 1.931 | 1.875 | 1.846 |
|  | $Ti_3C_2Cl_2$ | 3.459 | 2.826 | 2.666 | 2.602 | 2.357 | 2.459 |

Next, we explored the $Li_2Se_n$ - $Ti_3C_2S_2$ interactions, and the most stable relaxed configurations are shown in Figure Similar to $Ti_3C_2O_2$, in the case of $Ti_3C_2S_2$, the $Se_8$ molecule remains parallel to the surface with the shortest bond distance between the Se-S atoms as 3.256 Å and the value is greater when compared to $Ti_3C_2O_2$. The $Se_8$ adsorption on the $Ti_3C_2S_2$ retains the geometric configuration, however, the deformations are observed for the polyselenides. Likewise, $Ti_3C_2O_2$, from Figure 2, we observe that the binding energies for $Li_2Se_n$ on $Ti_3C_2S_2$ are in the order of 1.12 eV for $Se_8$, 1.41 eV for $Li_2Se_8$, 1.31 eV for $Li_2Se_6$, 1.69 eV for $Li_2Se_4$, 3.63 eV for $Li_2Se_2$, and 3.30 eV for $Li_2Se$. The binding energy data indicates that the $Li_2Se$ possesses the highest binding energy while $Li_2Se_6$ being the lowest among other $Li_2Se_n$ species. For both $Ti_3C_2O_2$ and $Ti_3C_2S_2$, the $Se_8$ exhibits the lowest binding energy.

Similarly, the adsorption energies for $Ti_3C_2F_2$ and $Ti_3C_2Cl_2$ are calculated from the optimized geometric configurations (Figure S6 and S7) and displayed in Figure 2. However, the adsorption energy values of $Se_8$ are found to be larger when compared to other higher order polyselenides indicating weaker $Li_2Se_n$ adsorption behavior that could result in ineffective anchoring. With the lithiation, both the class of materials exhibit a decrease in the adsorption energy values for $Li_2Se_6$, and the value increases as the lithiation further proceeds. The shortest bond distance between $Se_8$, $Li_2Se_n$, and AMs are calculated from energetically the most favorable configurations and are tabulated in Table 1. From the overall adsorption energy data of $Ti_3C_2F_2$ and $Ti_3C_2Cl_2$, we observe the poor adsorption behavior when compared to other functionalized $Ti_3C_2X_2$.

Next, in order to assess the anchoring effectiveness of the functionalized $Ti_3C_2$ MXenes— to provide adequate binding strength for preventing dissolution of higher order polyselenides ($Li_2Se_n \geq 4$) into the electrolyte—we calculated binding strength of $Li_2Se_n \geq 4$ with the commonly used 1,3 – dioxolane (DOL) and 1,2 dimethoxymethane (DME) electrolyte solvent molecules and compared them with the corresponding binding energies with the surface functionalized MXenes and graphene. Typically, the binding energies of polychalcogenides with the solvent molecules are calculated considering the interaction of one Li atom of polychalcogenide through a Li–O covalent bond.[36,46] The binding energies of $Li_2Se_n \geq 4$ bonded with the solvent molecules are represented in Figure 2, and the most favorable configurations are displayed in Figure S8. It is reported that the stronger interactions between the higher order polyselenides and the AMs than that of the solvent molecules is a required condition to prevent polyselenides dissolution into the electrolytes.[29,36] Figure 2 clearly indicates that out of all the materials considered, only $Ti_3C_2S_2$ and $Ti_3C_2O_2$ exhibit higher binding energies for $Li_2Se_n \geq 4$ compared to the DOL/DME cases. Therefore, it can be concluded that the higher thermodynamic stability of $Li_2Se_n \geq 4$ adsorbed on $Ti_3C_2S_2$ and $Ti_3C_2O_2$ structures can effectively constrain the higher order polyselenides from the dissolution into the electrolytes.

Next, we analyze the geometric features of the $Li_2Se_n$ adsorbed AMs. From Table 1, one can see that for all the cases of graphene and $Ti_3C_2X_2$, the largest distance between the AMs and adsorbate are found for the $Se_8$, and Figure 3 and S4-7 shows no structural deformation of the molecular. In the cases of $Li_2Se_n$, the Li atoms are found closer to the AMs than the Se atoms, and the shortest distance between the Li in $Li_2Se_n$ and the surfaces increases with the increasing count of the Se atoms, and this is predominantly because of the electronegative nature of Se and stronger chemical interactions. The trend of stronger $Li_2Se_n$ and AMs interactions with the decreasing Se count is correlated with the weakening of the intermolecular Li-Se bond strengths, which is evident from the increasing Li-Se bond lengths in $Li_2Se_n$.

Finally, observing the polyselenides adsorption energies of various functionalized $Ti_3C_2X_2$, we conclude that both S and O functionalization results in superior anchoring behavior compared to F and Cl functionalization and graphene. The moderate polyselenides binding energies of O and S functionalization can be deemed as adequate for the containment of $Li_2Se_n$ within the Se-cathode material without chemical decompositions.

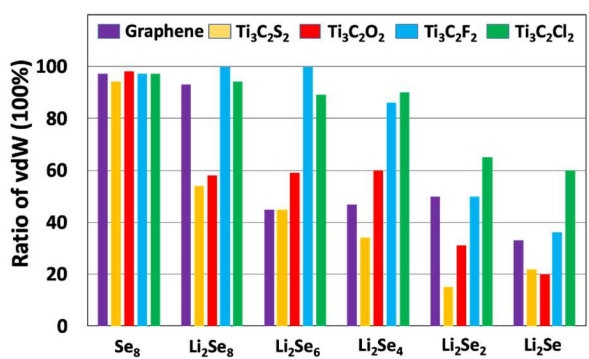

Figure 4. The ratio of vdW interaction of $Se_8$ and $Li_2Se_n$ adsorbed on Graphene and $Ti_3C_2X_2$

Furthermore, we probed the relative strength of the chemical and van der Waal interactions employing the ratio of vdW calculated using the formula, $R = \frac{\left(E_b^{vdw} - E_b^{novdw}\right)}{E_b^{vdw}}$ where $E_b^{vdw}$ and $E_b^{novdw}$ are the binding energies of $Li_2Se_n$ with and without van der Waal interactions, respectively. The ratio of vdW interaction for graphene and various AMs are shown in Figure 4. From the figure, we could observe that for all the AMs, the ratio of vdW interaction is nearly 100 % in the case of $Se_8$, which clearly demonstrates that the $Se_8$ interactions with the AMs are predominantly via non-bonded vdW. However, as the lithiation process continues, the ratio of vdW drops off, and the trend is found the opposite to the binding energy values. The interaction of Li atoms with the AMs results in stronger chemical interactions, which is evident from the reduction of the vdW ratio, and the highest chemical interactions are found in the $Li_2Se$ cases. For $Ti_3C_2S_2$ and $Ti_3C_2O_2$, we observe a systematic decrease in the vdW interactions with lithiation, however, in cases of $Ti_3C_2F_2$ and $Ti_3C_2Cl_2$, the ratio of vdW is greater than 50 percent except for $Ti_3C_2F_2$ – $Li_2Se$, which indicates that the lithiation process is dominated by the vdW interaction and these values are well consistent with other work.[29]

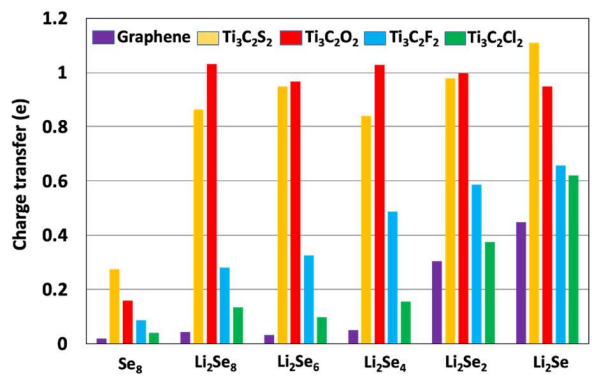

Figure 5. Charge transfer between the $Se_8$, polyselenides and graphene, $Ti_3C_2X_2$

***Bader Charge Analysis.*** In order to obtain further insights into the $Li_2Se_n$ binding mechanisms, Bader charge and charge density difference analysis were performed for the graphene and different surface terminal groups. The charge transfer values are displayed in Figure 5. From the figure, it is evident that for all the $Ti_3C_2X_2$, with the discharge process, electron transfer from the polyselenides to the $Ti_3C_2X_2$ increases indicating stronger chemical interactions between them. However, in the case of graphene, the charge transfer values are very minimal, indicating the weaker adsorption of the $Li_2Se_n$ on the surface because of the apolar characteristics of graphene. All the materials depict a similar trend in the charge transfer values, that is, the values increase from higher to lower order polyselenides except $Li_2Se_6$ species. The lower charge transfer value of $Li_2Se_6$ is correlated with the lower adsorption energies. We observe that the amount of charge transfer in $Ti_3C_2S_2$ and $Ti_3C_2O_2$ are greater when compared to F and Cl functionalization and graphene, which demonstrates stronger chemical interactions of $Li_2Se_n$ with S and O terminated surfaces. However, the lower charge transfer values for F and Cl cases can be attributed to the weaker chemical interaction between the $Li_2Se_n$ and the AMs. The calculated charge transfer values for $Ti_3C_2S_2$ adsorbed structures are 0.274 |e| for $Se_8$, 0.864 |e| for $Li_2Se_8$, 0.949 |e| for $Li_2Se_6$, 0.840 |e| for $Li_2Se_4$, 0.980 |e| for $Li_2Se_2$ and 1.110 |e| for $Li_2Se$. For $Ti_3C_2O_2$, the values are 0.159 |e| for $Se_8$, 1.031 |e| for $Li_2Se_8$, 0.967 |e| for $Li_2Se_6$, 1.027 |e| for $Li_2Se_4$, 0.998 |e| for $Li_2Se_2$ and 0.949 |e| for $Li_2Se$, respectively. The positive values for the charge transfer values observed in all the $Ti_3C_2X_2$ materials indicate that the charge is transferred from the $Li_2Se_n$ species to the $Ti_3C_2X_2$. The results illustrate that the charge transfer value of $Se_8$ is the lowest for all the functionalized groups when compared to other $Li_2Se_n$. The softening of the adsorbed intramolecular Li-Se bonds can be explained by considering the charge transfer process from the polyselenides to the $Ti_3C_2X_2$. The electrons are bereaved from the $Li_2Se_n$ as they are absorbed on the $Ti_3C_2X_2$. The higher electron deficiency in the $Li_2Se_n$ at the adsorbed state—results in enhanced chemical interactions of Li with the AMs—causes larger elongation of Li-Se bonds in $Li_2Se_n$. The charge transfer is primarily contributed from the Li atom to the AMs because of the lesser differences in the electronegativity between Se and surface atoms and the distance of Se atoms from the surface. To visualize the characteristics of charge transfer, we performed differential charge density analysis on representative cases such as $Li_2Se$, $Li_2Se_4$, and $Li_2Se_8$ and are displayed in Figure 6, S9, and S10. The charge depletion and accumulations observed from the analysis are in accordance with the quantity of the charge transfer values obtained from the Bader charge analysis.

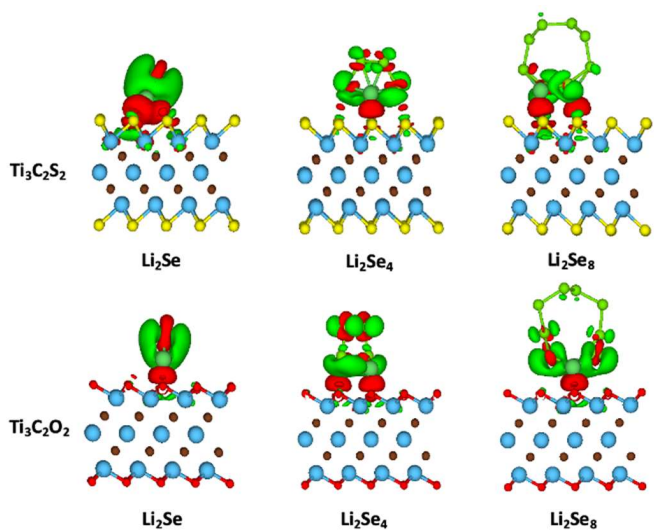

Figure 6. Side views of charge density difference of Li$_2$Se, Li$_2$Se$_4$ and Li$_2$Se$_8$ on Ti$_3$C$_2$S$_2$ and Ti$_3$C$_2$O$_2$. The iso-surface level is set at 0.0020 e Å$^{-3}$. The green and red colors denote charge accumulation and depletion, respectively.

***Density of States (DOS).*** We further studied the bonding characteristics between the Li$_2$Se$_n$ and the Ti$_3$C$_2$X$_2$ by examining the electronic structures of the adsorbed systems. We calculated the total and projected DOS (PDOS) to probe the conductive behavior of the Ti$_3$C$_2$X$_2$. The total DOS of pristine Ti$_3$C$_2$X$_2$, Li$_2$Se, Li$_2$Se$_4$, and Li$_2$Se$_8$ adsorbed on Ti$_3$C$_2$X$_2$ systems (Fig S10-14) were performed as representative cases of higher, medium, and lower order polyselenides to understand the change in electronic structure due to the charge transfer from the Li$_2$Se$_n$ to the functionalized MXenes before and after the adsorption. However, in all Li$_2$Se$_n$ cases, there is no significant change in the DOS for all the functional groups regardless of the Li$_2$Se$_n$ adsorption. From the figure S10-S14, we could observe that, compared to the pristine AMs, the Li$_2$Se$_n$ adsorbed systems results in negligible changes in the electronic structures as such electronic conductivity of the AMs are retained even after the Li$_2$Se$_n$ adsorption. The electronic conductive properties facilitate electron transfer directly to the involved redox reactions of Li$_2$Se$_n$ intermediates. In order to further investigate the influence of Li$_2$Se$_n$ on the Ti$_3$C$_2$X$_2$, projected PDOS was analyzed for both pristine and adsorbed systems of Ti$_3$C$_2$S$_2$ and Ti$_3$C$_2$O$_2$ and displayed in Figure 7. From the figure, we could observe that compared to pristine AMs, a considerable amount of electron states crosses the Fermi level in the case of Li$_2$Se$_n$ adsorbed on Ti$_3$C$_2$S$_2$ and Ti$_3$C$_2$O$_2$ indicates that the host materials could effectively circumvent the insulating behavior of Se$_8$.

Moreover, the metallic behavior of Li$_2$Se$_n$ adsorbed Ti$_3$C$_2$S$_2$ and Ti$_3$C$_2$O$_2$ is maintained due to the abundance of electronic states in the fermi energy level stemming from the 3d state of Ti. Moreover, a strong or partial polar covalent bond is formed from the hybridization of the Ti-3d and 2p/3p of S/O, resulting in Ti-S/O bonds. Hence all the adsorbed systems possess good electrical conductivity owing to the contribution of electrons from the Ti atoms, which is essential to expedite electrochemical reactions in the Li- Se battery systems.

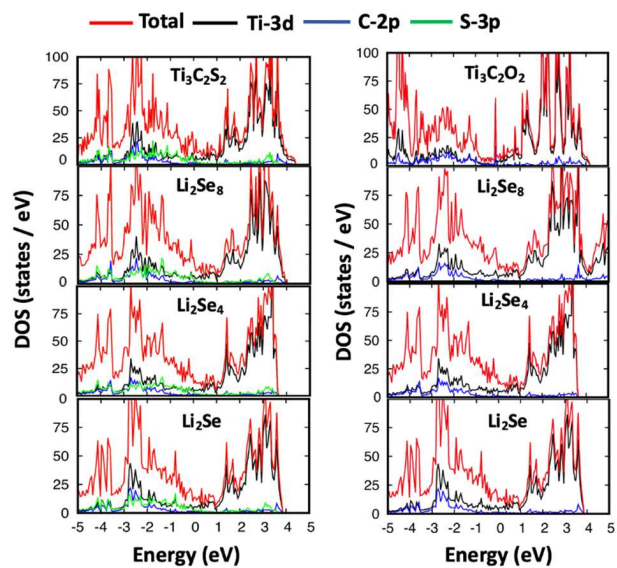

Fig 7. (a) Projected density of states of representative polyselenides cases for Ti$_3$C$_2$S$_2$ and (b) Ti$_3$C$_2$O$_2$

## 4. Conclusion

In summary, we employed the DFT calculations to demonstrate the functionality of graphene and Ti$_3$C$_2$ MXenes as host materials for Se-cathodes. We elucidate the mechanistic details of the inadequate polyselenides (Li$_2$Se$_n$) binding energies on graphene and strategies to improve the Li$_2$Se$_n$ retention capability within the Se-cathode using a new class of materials, Ti$_3$C$_2$ MXenes. The calculated weak adsorption energies resulted from Li$_2$Se$_n$ binding on graphene clearly indicate the graphene is incapable of providing requisite anchoring to prevent polyselenides from dissolution. Furthermore, we found that the bare Ti$_3$C$_2$ cannot be used as AM because of the stronger binding and subsequent dissociations of the polyselenides on the surface, which can debilitate the reversibility of the battery. We leveraged the modified Ti$_3$C$_2$ surfaces with various functional groups to overcome the limitations of bare Ti$_3$C$_2$. We identified Ti$_3$C$_2$S$_2$ and Ti$_3$C$_2$O$_2$ as AMs with superior performance, to provide adequate binding to Li$_2$Se$_n$ without chemical decomposition, compared to graphene and other functionalized MXenes. We consider the S- and

O- functionalized Ti$_3$C$_2$ as the appropriate choice for improved performance of the Se-based cathode for the Li$_2$Se$_n$ adsorption. The computed higher binding energies of polyselenides with Ti$_3$C$_2$S$_2$ and Ti$_3$C$_2$O$_2$ than the commonly used ether-based electrolyte solvents are indicative to the effectiveness to suppress the Li$_2$Se$_n$ migration. From the ratio of vdW analysis, we observed that the chemical interaction primarily originated from the intercalated Li$^+$ ions, and the van der Waal interactions were contributed by S. From the charge transfer analysis, we found that the charge is transferred from the Li$_2$Se$_n$ structure to the Ti$_3$C$_2$X$_2$. The density of states of Ti$_3$C$_2$X$_2$ reveals that the metallic properties of all the materials are retained even after adsorption with the polyselenides which is beneficial to promote the electrochemical process. Overall, the S- and O- functionalized Ti$_3$C$_2$ can be used to achieve two-fold objectives of (i) improving performance for polyselenides anchoring and (ii) facilitating the electronic conduction path to overcome the limitations of the insulating nature of Se as well as to eliminate the use of carbons. Thus, the utilization of the Ti$_3$C$_2$S$_2$ and Ti$_3$C$_2$O$_2$ MXenes in Se-based cathodes can be deemed as a sizeable step forward towards designing electrochemically inert carbon-free high-performance Se-cathode materials.

## Acknowledgments

M.M.I acknowledges the start-up funds from Wayne State University and the support of the Extreme Science and Engineering Discovery Environment (XSEDE)[47] for providing the computational facilities (Start-up Allocation – DMR190089).

## References

1  Armand, M.; Tarascon, J.-M. Building Better Batteries. *Nature* **2008**, *451* (7179), 652–657.
2  Bruce, P. G.; Freunberger, S. A.; Hardwick, L. J.; Tarascon, J.-M. Li-O2 and Li-S Batteries with High Energy Storage. *Nat. Mater.* **2011**, *11* (1), 19–29.
3  Ji, X.; Nazar, L. F. Advances in Li–S Batteries. *J. Mater. Chem.* **2010**, *20* (44), 9821–9826.
4  Nazar, L. F.; Cuisinier, M.; Pang, Q. Lithium-Sulfur Batteries. *MRS Bull.* **2014**, *39* (05), 436–442.
5  Chung, S.-Y.; Bloking, J. T.; Chiang, Y.-M. Electronically Conductive Phospho-Olivines as Lithium Storage Electrodes. *Nat. Mater.* **2002**, *1* (2), 123–128.
6  Ji, X.; Lee, K. T.; Nazar, L. F. A Highly Ordered Nanostructured Carbon-Sulphur Cathode for Lithium-Sulphur Batteries. *Nat. Mater.* **2009**, *8* (6), 500–506.
7  Liang, X.; Wen, Z.; Liu, Y.; Wu, M.; Jin, J.; Zhang, H.; Wu, X. Improved Cycling Performances of Lithium Sulfur Batteries with LiNO<SUB>3</SUB>-Modified Electrolyte. *J. Power Sources* **2011**, *196*, 9839.
8  Rezan, D. *Li-s Batteries: The Challenges, Chemistry, Materials, And Future Perspectives*; #N/A, 2017.
9  Abouimrane, A.; Dambournet, D.; Chapman, K. W.; Chupas, P. J.; Weng, W.; Amine, K. A New Class of Lithium and Sodium Rechargeable Batteries Based on Selenium and Selenium–Sulfur as a Positive Electrode. *J. Am. Chem. Soc.* **2012**, *134* (10), 4505–4508.
10 Liu, L.; Hou, Y.; Wu, X.; Xiao, S.; Chang, Z.; Yang, Y.; Wu, Y. Nanoporous Selenium as a Cathode Material for Rechargeable Lithium–Selenium Batteries. *Chem. Commun.* **2013**, *49* (98), 11515–11517.
11 Cai, Q.; Li, Y.; wang, L.; Li, Q.; Xu, J.; Gao, B.; Zhang, X.; Huo, K.; Chu, P. K. Freestanding Hollow Double-Shell Se@CNx Nanobelts as Large-Capacity and High-Rate Cathodes for Li-Se Batteries. *Nano Energy* **2017**, *32*, 1–9.
12 Yang, J.; Gao, H.; Ma, D.; Zou, J.; Lin, Z.; Kang, X.; Chen, S. High-Performance Li-Se Battery Cathode Based on CoSe2-Porous Carbon Composites. *Electrochimica Acta* **2018**, *264*, 341–349.
13 Eftekhari, A. The Rise of Lithium–Selenium Batteries. *Sustain. Energy Fuels* **2017**, *1* (1), 14–29.
14 Yang, C.-P.; Yin, Y.-X.; Guo, Y.-G. Elemental Selenium for Electrochemical Energy Storage. *J. Phys. Chem. Lett.* **2015**, *6* (2), 256–266.
15 Li, Z.; Yin, L. MOF-Derived, N-Doped, Hierarchically Porous Carbon Sponges as Immobilizers to Confine Selenium as Cathodes for Li–Se Batteries with Superior Storage Capacity and Perfect Cycling Stability. *Nanoscale* **2015**, *7* (21), 9597–9606.
16 Yuan, W.; Feng, Y.; Xie, A.; Zhang, X.; Huang, F.; Li, S.; Zhang, X.; Shen, Y. Nitrogen-Doped Nanoporous Carbon Derived from Waste Pomelo Peel as a Metal-Free Electrocatalyst for the Oxygen Reduction Reaction. *Nanoscale* **2016**, *8* (16), 8704–8711.
17 Luo, C.; Xu, Y.; Zhu, Y.; Liu, Y.; Zheng, S.; Liu, Y.; Langrock, A.; Wang, C. Selenium@Mesoporous Carbon Composite with Superior Lithium and Sodium Storage Capacity. *ACS Nano* **2013**, *7* (9), 8003–8010.
18 Liu, T.; Dai, C.; Jia, M.; Liu, D.; Bao, S.; Jiang, J.; Xu, M.; Li, C. M. Selenium Embedded in Metal-Organic Framework Derived Hollow Hierarchical Porous Carbon Spheres for Advanced Lithium-Selenium Batteries. *ACS Appl. Mater. Interfaces* **2016**, *8* (25), 16063–16070.
19 Zhang, S.-F.; Wang, W.-P.; Xin, S.; Ye, H.; Yin, Y.-X.; Guo, Y.-G. Graphitic Nanocarbon–Selenium Cathode with Favorable Rate Capability for Li–Se Batteries. *ACS Appl. Mater. Interfaces* **2017**, *9* (10), 8759–8765.
20 Liu, T.; Zhang, Y.; Hou, J.; Lu, S.; Jiang, J.; Xu, M. High Performance Mesoporous C@Se Composite Cathodes Derived from Ni-Based MOFs for Li–Se Batteries. *RSC Adv.* **2015**, *5* (102), 84038–84043.
21 Strongly Bonded Selenium/Microporous Carbon Nanofibers Composite as a High-Performance Cathode for Lithium–Selenium Batteries | The Journal of Physical Chemistry C https://pubs.acs.org/doi/abs/10.1021/acs.jpcc.5b09553 (accessed Nov 25, 2019).
22 Zhang, J.; Zhang, Z.; Li, Q.; Qu, Y.; Jiang, S. Selenium Encapsulated into Interconnected Polymer-Derived Porous Carbon Nanofiber Webs as Cathode Materials for Lithium-Selenium Batteries. *J. Electrochem. Soc.* **2014**, *161* (14), A2093–A2098.
23 Youn, H.-C.; Jeong, J. H.; Roh, K. C.; Kim, K.-B. Graphene–Selenium Hybrid Microballs as Cathode Materials for High-Performance Lithium–Selenium Secondary Battery Applications. *Sci. Rep.* **2016**, *6* (1), 1–8.
24 Zeng, L.; Zeng, W.; Jiang, Y.; Wei, X.; Li, W.; Yang, C.; Zhu, Y.; Yu, Y. A Flexible Porous Carbon Nanofibers-Selenium Cathode with Superior Electrochemical Performance for Both Li-Se and Na-Se Batteries. *Adv. Energy Mater.* **2015**, *5* (4), 1401377.
25 Yi, Z.; Yuan, L.; Sun, D.; Li, Z.; Wu, C.; Yang, W.; Wen, Y.; Shan, B.; Huang, Y. High-Performance Lithium–Selenium Batteries Promoted by Heteroatom-Doped Microporous Carbon. *J. Mater. Chem. A* **2015**, *3* (6), 3059–3065.
26 Han, K.; Liu, Z.; Shen, J.; Lin, Y.; Dai, F.; Ye, H. A Free-Standing and Ultralong-Life Lithium-Selenium Battery Cathode Enabled by 3D Mesoporous Carbon/Graphene Hierarchical Architecture. *Adv. Funct. Mater.* **2015**, *25* (3), 455–463.
27 He, J.; Chen, Y.; Lv, W.; Wen, K.; Li, P.; Wang, Z.; Zhang, W.; Qin, W.; He, W. Three-Dimensional Hierarchical Graphene-CNT@Se: A Highly Efficient Freestanding Cathode for Li–Se Batteries. *ACS Energy Lett.* **2016**, *1* (1), 16–20.
28 Mi, K.; Jiang, Y.; Feng, J.; Qian, Y.; Xiong, S. Hierarchical Carbon Nanotubes with a Thick Microporous Wall and Inner Channel as Efficient Scaffolds for Lithium–Sulfur Batteries. *Adv. Funct. Mater.* **2016**, *26* (10), 1571–1579.
29 Wang, D.; Li, F.; Lian, R.; Xu, J.; Kan, D.; Liu, Y.; Chen, G.; Gogotsi, Y.; Wei, Y. A General Atomic Surface Modification Strategy for Improving

Anchoring and Electrocatalysis Behavior of Ti3C2T2 MXene in Lithium-Sulfur Batteries. *ACS Nano* **2019**, *13* (10), 11078–11086.
30  Naguib, M.; Kurtoglu, M.; Presser, V.; Lu, J.; Niu, J.; Heon, M.; Hultman, L.; Gogotsi, Y.; Barsoum, M. W. Two-Dimensional Nanocrystals Produced by Exfoliation of Ti3AlC2. *Adv. Mater.* **2011**, *23* (37), 4248–4253.
31  Hu, M.; Li, Z.; Hu, T.; Zhu, S.; Zhang, C.; Wang, X. High-Capacitance Mechanism for Ti3C2Tx MXene by in Situ Electrochemical Raman Spectroscopy Investigation. *ACS Nano* **2016**, *10* (12), 11344–11350.
32  Liang, X.; Rangom, Y.; Kwok, C. Y.; Pang, Q.; Nazar, L. F. Interwoven MXene Nanosheet/Carbon-Nanotube Composites as Li-S Cathode Hosts. *Adv. Mater. Deerfield Beach Fla* **2017**, *29* (3).
33  Hu, J.; Xu, B.; Ouyang, C.; Yang, S. A.; Yao, Y. Investigations on V2C and V2CX2 (X = F, OH) Monolayer as a Promising Anode Material for Li Ion Batteries from First-Principles Calculations. *J. Phys. Chem. C* **2014**, *118* (42), 24274–24281.
34  Are MXenes Promising Anode Materials for Li Ion Batteries? Computational Studies on Electronic Properties and Li Storage Capability of Ti3C2 and Ti3C2X2 (X = F, OH) Monolayer | Journal of the American Chemical Society https://pubs.acs.org/doi/10.1021/ja308463r (accessed Nov 25, 2019).
35  Yu, Y.-X. Prediction of Mobility, Enhanced Storage Capacity, and Volume Change during Sodiation on Interlayer-Expanded Functionalized Ti3C2 MXene Anode Materials for Sodium-Ion Batteries. *J. Phys. Chem. C* **2016**, *120* (10), 5288–5296.
36  Wang, Y.; Shen, J.; Xu, L.-C.; Yang, Z.; Li, R.; Liu, R.; Li, X. Sulfur-Functionalized Vanadium Carbide MXene (V2CS2) as a Promising Anchoring Material for Lithium–Sulfur Batteries. *Phys. Chem. Chem. Phys.* **2019**, *21* (34), 18559–18568.
37  Kresse, G.; Furthmüller, J. Efficient Iterative Schemes for Ab Initio Total-Energy Calculations Using a Plane-Wave Basis Set. *Phys. Rev. B* **1996**, *54* (16), 11169.
38  Perdew, J. P.; Burke, K.; Ernzerhof, M. Generalized Gradient Approximation Made Simple. *Phys. Rev. Lett.* **1996**, *77* (18), 3865–3868.
39  Grimme, S.; Antony, J.; Ehrlich, S.; Krieg, H. A Consistent and Accurate Ab Initio Parametrization of Density Functional Dispersion Correction (DFT-D) for the 94 Elements H-Pu. *J. Chem. Phys.* **2010**, *132* (15), 154104.
40  Grimme, S.; Ehrlich, S.; Goerigk, L. Effect of the Damping Function in Dispersion Corrected Density Functional Theory. *J. Comput. Chem.* **2011**, *32* (7), 1456–1465.
41  Henkelman, G.; Arnaldsson, A.; Jónsson, H. A Fast and Robust Algorithm for Bader Decomposition of Charge Density. *Comput. Mater. Sci.* **2006**, *36* (3), 354–360.
42  Momma, K.; Izumi, F. VESTA 3 for Three-Dimensional Visualization of Crystal, Volumetric and Morphology Data. *J. Appl. Crystallogr.* **2011**, *44* (6), 1272–1276.
43  Qiu, Y.; Li, W.; Zhao, W.; Li, G.; Hou, Y.; Liu, M.; Zhou, L.; Ye, F.; Li, H.; Wei, Z.; et al. High-Rate, Ultralong Cycle-Life Lithium/Sulfur Batteries Enabled by Nitrogen-Doped Graphene. *Nano Lett.* **2014**, *14* (8), 4821–4827.
44  Tang, Q.; Zhou, Z.; Shen, P. Are MXenes Promising Anode Materials for Li Ion Batteries? Computational Studies on Electronic Properties and Li Storage Capability of Ti3C2 and Ti3C2X2 (X = F, OH) Monolayer. *J. Am. Chem. Soc.* **2012**, *134* (40), 16909–16916.
45  Wang, H.-W.; Naguib, M.; Page, K.; Wesolowski, D. J.; Gogotsi, Y. Resolving the Structure of Ti3C2Tx MXenes through Multilevel Structural Modeling of the Atomic Pair Distribution Function. *Chem. Mater.* **2016**, *28* (1), 349–359.
46  Wang, B.; Alhassan, S. M.; Pantelides, S. T. Formation of Large Polysulfide Complexes during the Lithium-Sulfur Battery Discharge. *Phys. Rev. Appl.* **2014**, *2* (3), 034004.
47  Towns, J.; Cockerill, T.; Dahan, M.; Foster, I.; Gaither, K.; Grimshaw, A.; Hazlewood, V.; Lathrop, S.; Lifka, D.; Peterson, G. D. XSEDE: Accelerating Scientific Discovery. *Comput. Sci. Eng.* **2014**, *16* (5), 62–74.

## Supporting Information

Table S1. Structural data of the bare and functionalized Mxenes

| AM | Lattice Parameters (Å) | | Bond length (Å) | |
|---|---|---|---|---|
| | a = b | Ti – C | Ti – Ti | Ti – S/O/F/Cl |
| Ti$_3$C$_2$ | 12.370 | 2.05 | 2.92 | - |
| Ti$_3$C$_2$S$_2$ | 12.493 | 2.16 | 3.03 | 2.40 |
| Ti$_3$C$_2$O$_2$ | 12.108 | 2.18 | 3.10 | 1.97 |
| Ti$_3$C$_2$F$_2$ | 12.215 | 2.06 | 2.94 | 2.16 |
| Ti$_3$C$_2$Cl$_2$ | 12.108 | 2.05 | 2.93 | 2.47 |

Table S2. The binding energies (in eV) of higher order polyselenides with the electrolyte solvents

| Higher order Li$_2$Se$_n$ polyselenides | DME | DOL |
|---|---|---|
| Li$_2$Se$_4$ | 0.87 | 0.82 |
| Li$_2$Se$_6$ | 0.92 | 0.79 |
| Li$_2$Se$_8$ | 0.90 | 0.83 |

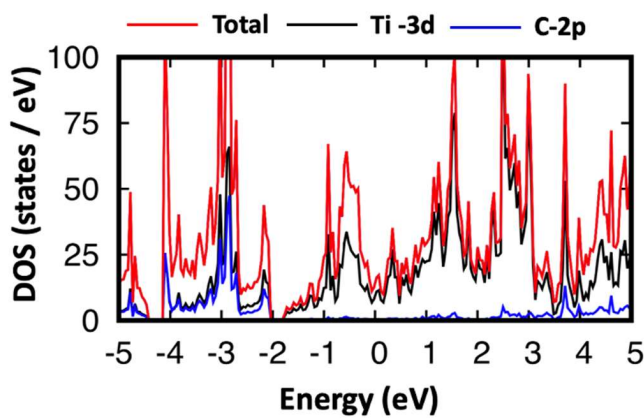

Fig S1. partial DOS of Ti3C2

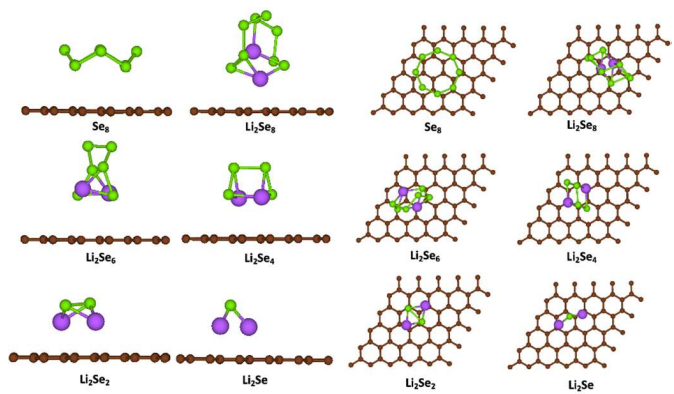

Figure S2. Top and side views of the structures for Li$_2$Se$_n$ adsorption on monolayer graphene

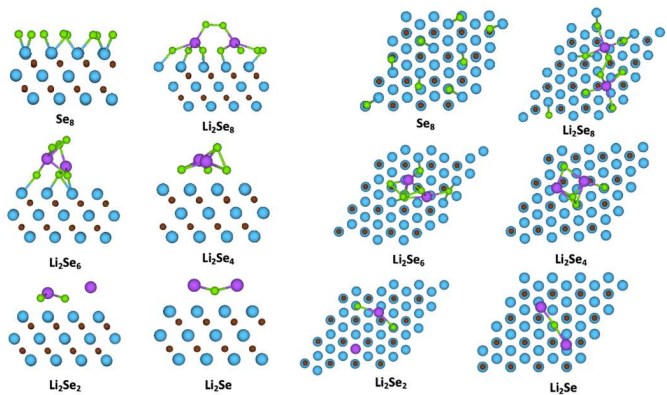

Figure S3. The side and top views of the most stable structures of Ti$_3$C$_2$

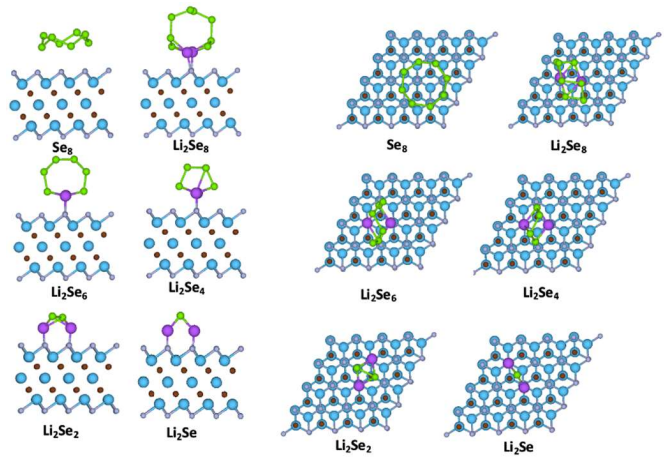

Figure S6. The side and top views of the optimized structures of Li$_2$Se$_n$ adsorbed Ti$_3$C$_2$F$_2$

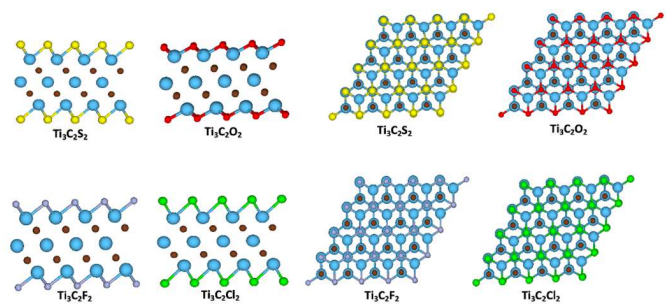

Figure S4. The side and top views of the optimized structures of Ti$_3$C$_2$X$_2$ (X = S, O, F, Cl)

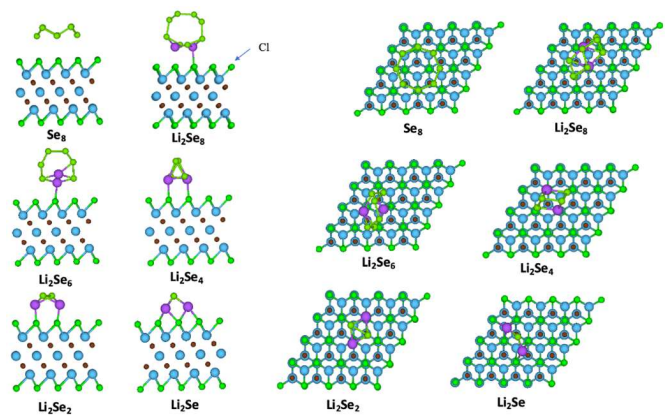

Figure S7. The side and top views of the most favorable sites of Li$_2$Se$_n$ adsorbed Ti$_3$C$_2$Cl$_2$

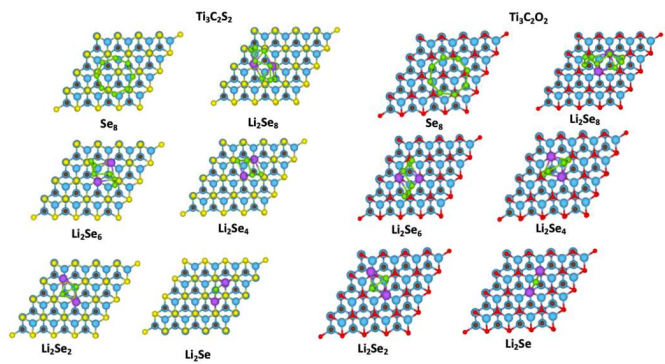

Figure S5. The top views of the most stable structures of Li$_2$Se$_n$ adsorbed on Ti$_3$C$_2$S$_2$ and Ti$_3$C$_2$O$_2$.

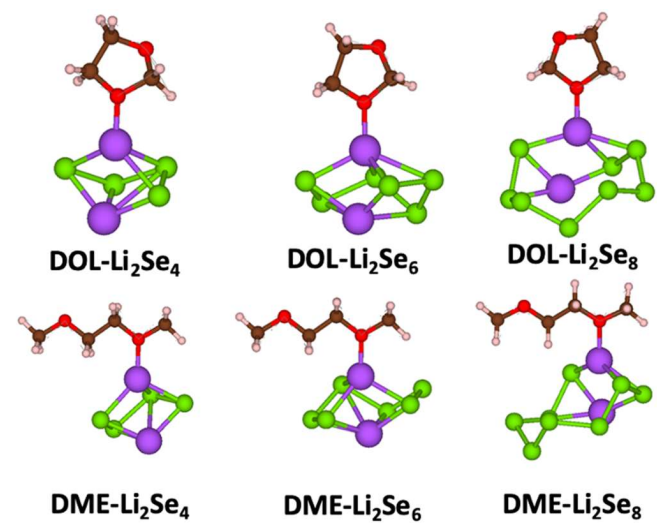

Figure S8. The most stable geometric configurations of (Li$_2$Se$_n$ ≥ 4) adsorbed DOL/ DME solvents

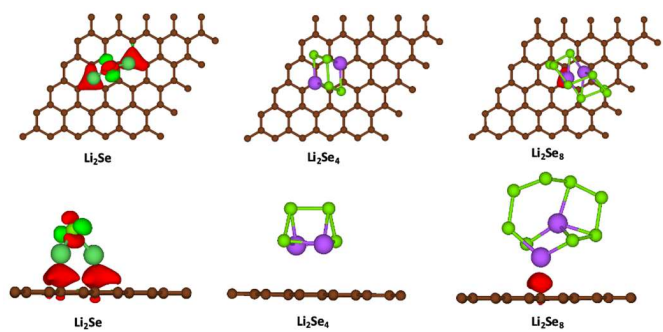

Figure S9. Charge density difference of Li$_2$Se$_n$ (n = 1,4,8) on Graphene

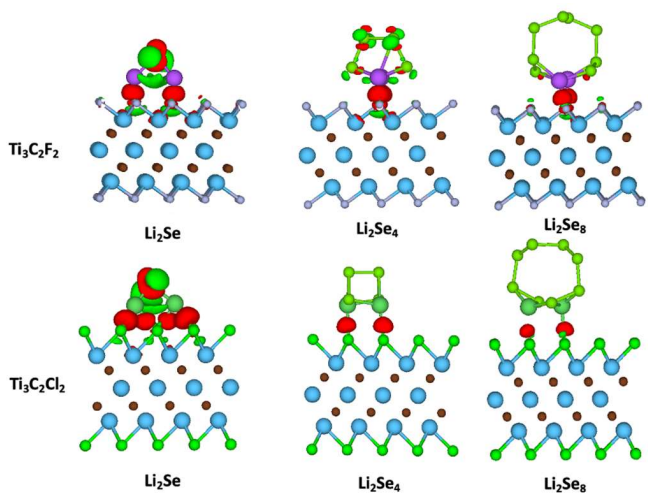

Figure S10. Charge density difference of Li$_2$Se$_n$ (n = 1,4,8) on Ti$_3$C$_2$F$_2$ and Ti$_3$C$_2$Cl$_2$

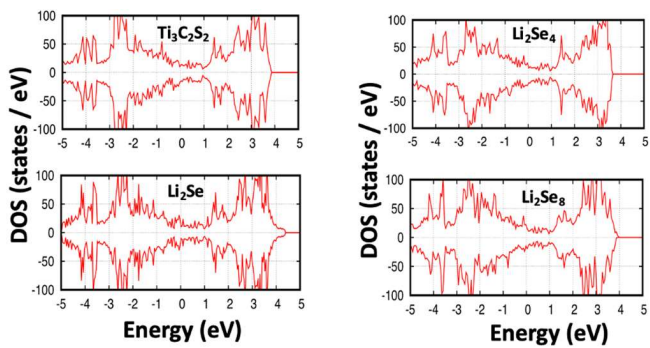
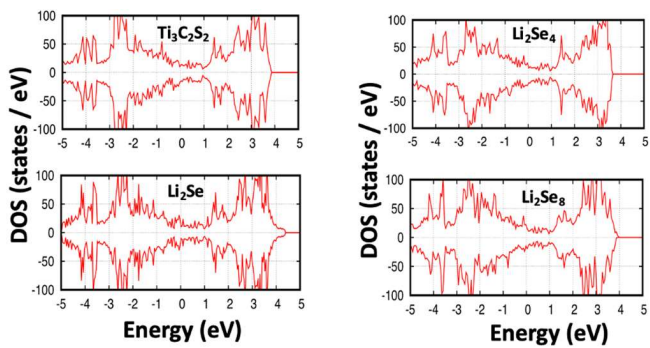

Figure S11. TDOS of Li$_2$Se$_n$ (n = 1,4,8) adsorbed Ti$_3$C$_2$S$_2$

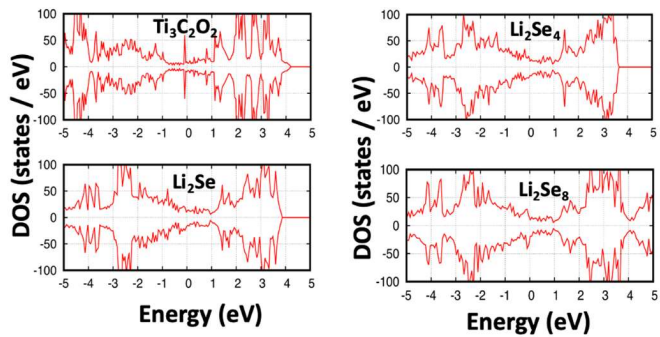

Figure S12. TDOS of Li$_2$Se$_n$ (n = 1,4,8) adsorbed Ti$_3$C$_2$O$_2$

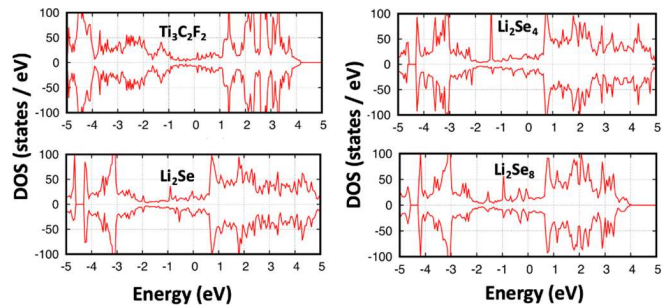

Figure S13. TDOS of Li$_2$Se$_n$ (n = 1,4,8) adsorbed Ti$_3$C$_2$F$_2$

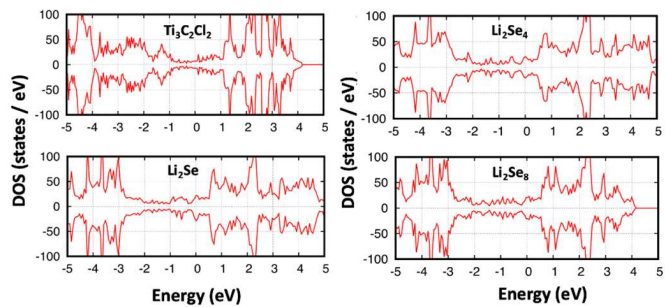

Figure S14. TDOS of Li$_2$Se$_n$ (n = 1,4,8) adsorbed Ti$_3$C$_2$Cl$_2$